\documentclass[
reprint,
superscriptaddress,
showpacs,preprintnumbers,
amsmath,amssymb,
aip
]{revtex4-1}

\usepackage{color}
\usepackage{graphicx}
\usepackage{dcolumn}
\usepackage{bm}
\usepackage{txfonts}
\usepackage{tikz}
\usepackage{pgffor}
\usepackage{verbatim}
\usepackage{pstricks}
\usepackage{pst-3dplot}
\usepackage[colorlinks, breaklinks=true, linkcolor=blue, citecolor=blue, linktocpage=true]{hyperref}

\begin{document}


\title{Effects of magnonic Kerr nonlinearity on magnon-polaritons with a soft-mode}

\author{Takahiro Chiba}
 \email{t.chiba@yz.yamagata-u.ac.jp}
 \affiliation{Department of Information Science and Technology, Graduate School of Science and Engineering, Yamagata University, Yonezawa, Yamagata 992-8510, Japan}
 \affiliation{Department of Applied Physics, Graduate School of Engineering, Tohoku University, Sendai, Miyagi 980-8579, Japan}


\date{\today}

\begin{abstract}
We theoretically study the effects of magnonic Kerr nonlinearity on magnon–polaritons (MPs) with a soft-mode in easy-axis ferromagnets coupled to a microwave cavity. Using an effective circuit model capable of describing MPs up to the nonperturbative strong-coupling regime, we show that chaotic and frequency-comb-like behaviors of MPs emerge at the original modes crossing point. Furthermore, we demonstrate that the Kerr nonlinearity induces a finite excitation gap in the soft-mode, particularly in the strong-coupling regime.
\end{abstract}

\maketitle

\section{Introduction}

Cavity magnonics, wherein the key excitation is a magnon-polariton (MP) that is a strongly coupled state of spin waves (magnons) and microwave photons, is one of the prominent platforms for novel quantum devices  \cite{Rameshti22,Yuan22,Li20}. In general, polariton states satisfying $g/\omega_{\rm c(m)} \gtrsim 0.3$ are referred to as the \textit{nonperturbative} strong-coupling (NSC) regime \cite{Diaz19,Kockum19}, where $\omega_{\rm c(m)}$ denotes the characteristic angular frequency of the cavity or matter excitation and $g$ is the coupling strength. In this regime, the excitation nonconserving nature of the light-matter interaction leads to strongly squeezed polariton states, giving rise to a nontrivial ground state characterized by virtual excitations and nonclassical correlations \cite{Baydin25}.

Very recently, NSC has been experimentally demonstrated in MP systems based on superconductor/ferromagnet nanostructures \cite{Golovchanskiy21SciAdv,Golovchanskiy21PRAP} and magnetic slab geometries \cite{Bourcin23}. Within cavity magnonics and linear spin-wave theory, perfect magnon squeezing has been predicted by exploiting soft-mode (zero-mode) magnons, whose excitation gap vanishes at the critical magnetic field \cite{Lee23,Silaev23,Bauer23}. Typically, such soft-modes arise in anisotropic magnets with canted magnetization configurations induced by an external magnetic field applied perpendicular to the anisotropy axis \cite{Iihama14,de Wal23}. These magnetic anisotropies give rise to nonlinear magnon effects, the simplest of which is the self-Kerr nonlinearity \cite{Zheng23}. Under sufficiently strong driving, the Kerr nonlinearity substantially modifies uniform magnon dynamics, leading to bistability \cite{Wang18}, frequency-comb generation \cite{Wang21}, and chaotic behavior \cite{Wang19,Elyasi20}.
Therefore, it is highly desirable to reveal the effects of magnonic Kerr nonlinearity on the dynamics of MPs in the NSC regime.

In this paper, we theoretically investigate how the Kerr nonlinearity affect the dynamics of MPs with a soft-mode in easy-axis ferromagnets. Using an effective circuit model capable of describing MPs up to the NSC regime  \cite{Chiba24APL,Chiba24MSJ,Mita25PRAP,Suzuki25,Chiba25}, we show that chaotic and frequency-comb-like behaviors of MP at the original modes crossing point in the typical strong coupling (SC) regime with $g/\omega_{\rm c(m)} \approx 0.01$. Furthermore, the Kerr nonlinearity induces a finite excitation gap in the soft-mode especially in the SC regime. In contrast, such nonlinear behaviors of MPs are found to be strongly suppressed in the NSC regime characterized by $g/\omega_{\rm c(m)} \approx 1$.

\section{Effective circuit model of magnon-polaritons} 

\begin{figure}[ptb]
\begin{centering}
\includegraphics[width=0.47\textwidth,angle=0]{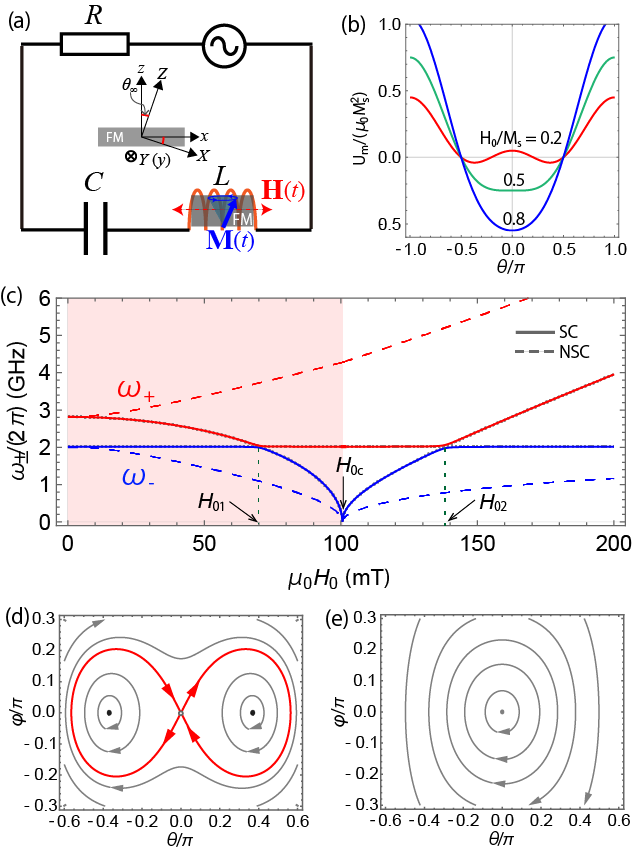} 
\par\end{centering}
\caption{
(a) Effective circuit model of a cavity magnonics system, where ${\bf M}(t)$ represents the uniform magnetization dynamics (magnons) in an easy-axis ferromagnet (FM) and ${\bf H}(t)$ denotes the microwave magnetic field (photons) in the inductor. The equilibrium magnetization is characterized by the angle $\theta_{\infty}$ measured from the $z$ axis.
(b) Total magnetic energy $U_{\rm m}$ as a function of the polar angle $\theta$ at $\varphi = 0$ for different values of $H_0/M_{\rm s}$.
(c) Eigenfrequencies ($\omega_\pm$) of MPs as a function of the external magnetic field ($\mu_0H_0$) for different ratios of $d_{\rm M}/d$: SC for $d_{\rm M}/d = 0.02$ and NSC for $d_{\rm M}/d = 1$. The critical field $H_{0c} = M_{\rm s}/2$ marks the transition of the equilibrium magnetization position.
(d),(e) Phase portraits for (b), where the red line, filled circles, and open circle denote the homoclinic orbit, centers, and saddle point, respectively: (d) $H_0 < H_{0c}$ and (e) $H_0 > H_{0c}$.
}
\label{Fig:system}
\end{figure}

Based on our previous works \cite{Chiba24APL,Chiba24MSJ,Mita25PRAP,Suzuki25,Chiba25}, we begin by reviewing the concept of an effective circuit model for cavity magnonics systems. As illustrated in Fig.~\ref{Fig:system}~(a), the effective circuit represents the microwave photon system as a single $LC$ resonator while the magnetization dynamics of an easy-axis ferromagnet inserted into the inductor correspond to the magnon system. Within the inductor, the magnetization of the ferromagnet is driven by the microwave magnetic field ${\bf H}(t)$ associated with the $LC$ resonator mode. The microwave photons thereby couple to the magnon dynamics, ${\bf M}(t)$, through the electromagnetic interaction. 
The magnon dynamics is described by the Landau-Lifshitz-Gilbert (LLG) equation
\begin{align}
\frac{d \bf M}{d t} = -\gamma{\bf M}\times\left(  -\frac{\delta U_{\rm m}}{\delta{\bf M}} +\mu_0k_{\rm c}{\bf H}(t)\right) + \frac{\alpha}{M_{\rm s}}{\bf M}\times\frac{d \bf M}{d t}
,\label{LLG}
\end{align}
where $M_{\rm s}$ is the saturation magnetization of a ferromagnet, $\gamma$ is the gyromagnetic ratio, $k_{\rm c}$ is the effective Nagaoka coefficient of an inductor, which is responsible for the mode volume of the microwave photon, $\mu_0$ is the permeability of free space, and $\alpha$ is the Gilbert damping constant. 
Here, the total magnetic energy is assumed to be
\begin{align}
U_{\rm m} = - \mu_0{\bf M}\cdot{\bf H}_0 - \frac{1}{2}\mu_0{\bf M}\cdot{\bf H}_{\rm d}[{\bf m}]
,\label{Um}
\end{align}
where ${\bf H}_0 = H_0\hat{\bf z}$ is the external static magnetic field and ${\bf H}_{\rm d}[{\bf m}] = - M_{\rm s}(  m_y\hat{\bf y} + m_z\hat{\bf z})/2$ describes the demagnetization field due to a cylindrical rod shape of a ferromagnet. Here, ${\bf m}(t) = {\bf M}(t)/M_{\rm s} = (m_x,m_y,m_z)$ is the unit vector along the magnetization direction of the ferromagnet.
Introducing the polar angle $\theta(t)$ and azimuthal angle $\varphi(t)$ for ${\bf m}(t) = (\cos\varphi\sin\theta, \sin\varphi\sin\theta, \cos\theta)$, Eq.~(\ref{Um}) can be expressed as
\begin{align}
\frac{U_{\rm m}}{\mu_0M_{\rm s}^2} = - \frac{H_0}{M_{\rm s}}\cos\theta + \frac{1}{4}\left( 1 - \cos^2\varphi\sin^2\theta\right)
,\label{Um2}
\end{align}
which gives the equilibrium position of ${\bf m}(t \to \infty)$ in the absence of ${\bf H}(t)$ by
$\left(\theta_{\infty}, \varphi_{\infty}\right) =  (\cos^{-1}(2H_0/M_{\rm s}), 0)$.
Note that a strong field, $H_0 \geq M_{\rm s}/2$, aligns the magnetization along the $z$ direction, resulting in $\theta_{\infty} = 0$ whereas a weaker field, $H_0 < M_{\rm s}/2$, produces canted magnetization configurations with two degenerate equilibrium positions at $\varphi_{\infty} = 0$ [see Fig.~\ref{Fig:system}~(b)].
In the $XYZ$-coordinate system depicted in Fig.~\ref{Fig:system}~(a), the magnetization is stabilized along the $Z$-axis determined by $\theta_{\infty}$ and $\varphi_{\infty}$. Denoting the transformation matrix as $\mathcal{R}(\theta_{\infty})$ around the $Y(y)$-axis, in the presence of ${\bf H}(t)$, the magnetization ${\bf n}(t) = \mathcal{R}(\theta_{\infty}){\bf m}(t) = (n_X,n_Y,n_Z)$ precesses around $\mu_0{\bf H}_{\mathcal{R}} = \mathcal{R}(\theta_{\infty})\left(  -\delta U_{\rm m}/\delta{\bf M}\right)$.
For a small microwave field ${\bf p}(t) = \mathcal{R}(\theta_{\infty}){\bf h}(t)$ given by ${\bf h}(t) = {\bf H}(t)/M_{\rm s}$, the magnetization dynamics can be linearized as ${\bf n}(t) = (n_X(t),n_Y(t),1)$ with $|n_X|,|n_Y| \ll 1$. 
Therefore, the linearized LLG equation gives the magnon eigenfrequncey
\begin{align}
\omega_{\rm m} = 
\begin{cases}
 \gamma'\mu_0\sqrt{\left(  M_{\rm s}/2\right)^2 - H_0^2} & (H_0 < M_{\rm s}/2) \\
 \gamma'\mu_0\sqrt{H_0\left(  H_0 - M_{\rm s}/2\right)} & (H_0 \geq M_{\rm s}/2)
\end{cases}
,\label{omegamrod}
\end{align}
where $\gamma' = \gamma/\sqrt{1 + \alpha^2}$. At $H_0 = M_{\rm s}/2$, the magnon eigenfrequncey once becomes zero, indicating the presence of soft magnons \cite{Bauer23}. 

Next, we investigate eigenmodes of MP in the framework of the effective circuit model.
Inside the inductor, the microwave field is induced by an alternating current $I(t)$ via Amp\`{e}re's law: ${\bf H}(t) = (NI(t)/l)\hat{\bf x}$, where $N$ is a turn number and $l$ is the inductor length. 
Defining the circuit mode angular frequency by $\omega_c = 1/\sqrt{LC}$ with an inductance $L$ and an electrostatic capacitance $C$, the dynamics of the microwave field is govern by \cite{Suzuki25,Chiba25}
\begin{align}
\left(  \frac{d^2}{dt^2} + 2\beta\omega_{\rm c}\frac{d}{dt} + \omega_{\rm c}^2\right){\bf H} + k_{\rm m}\eta_{\rm S}\frac{d^2{\bf M}}{dt^2}
= \omega_{\rm c}^2{\bf H}_V(t)
,\label{Maxwell}
\end{align}
where $\beta = R\sqrt{C/L}/2$ is an effective damping of the circuit with an electrical resistance $R$, $k_{\rm m}$ is the effective Nagaoka coefficient of a ferromagnet, $\eta_{\rm S}$~($0\leq\eta_{\rm S}\leq1$) represents the volume ratio between the inductor and ferromagnet, and ${\bf H}_V(t)$~(${\bf p}_V(t) = \mathcal{R}(\theta_{\infty}){\bf H}_V(t)/M_{\rm s}$ in the $XYZ$-coordinate) is an driving magnetic field from the source.
Assuming harmonic solutions $p_X(t) = \tilde{p}_Xe^{-i\omega t}$ and $n_X(t) = \tilde{n}_Xe^{-i\omega t}$ for the input $p_V(t) = p_Ve^{-i\omega t}$, we obtain the coupled equations in the frequency domain\cite{Suzuki25,Chiba25}
\begin{align}
\bar{\Omega}
\begin{pmatrix}
\tilde{p}_X \\
\tilde{n}_X
\end{pmatrix}
=
\begin{pmatrix}
-\omega_{\rm c}^2p_V \\
-\omega_{\parallel1}\omega_{\perp}
\end{pmatrix}
\label{LLG-RLC}
\end{align}
with 
\begin{align}
\bar{\Omega} = 
\begin{pmatrix}
\omega^2 + 2i \beta\omega_{\rm c}\omega - \omega_{\rm c}^2
& k_{\rm m}\eta_{\rm S}\cos^2\theta_{\infty}\omega^2 \\
k_{\rm c}\gamma'\mu_0M_{\rm s}\omega_{\parallel1}
& \omega^2
+ 2i\alpha_{\rm m}\omega_{\rm m}\omega 
- \omega_{\rm m}^2
\end{pmatrix}
,\label{OmegaMP}
\end{align}
where $\alpha_{\rm m}
= \alpha\left(  \omega_{\parallel1} + \omega_{\parallel2}\right)/2/\sqrt{1 + \alpha^2}/\omega_{\rm m}$,  
$\omega_{\parallel1}
= \gamma'\mu_0\left(  H_0\cos\theta_{\infty} 
+ M_{\rm s}\sin^2\theta_{\infty}/2\right)$, 
$\omega_{\parallel2}
= \gamma'\mu_0\left(  H_0\cos\theta_{\infty} 
- M_{\rm s}\cos2\theta_{\infty}/2\right)$,
and
$\omega_{\perp}
= \gamma'\mu_0\left(  -H_0\sin\theta_{\infty}
+ M_{\rm s}\sin2\theta_{\infty}/4\right)$.
In order to obtain the eigenfrequency of the hybridized MP modes, we neglect the damping parameters in Eq.~(\ref{OmegaMP}) by setting $\alpha_{\rm m} = \beta = 0$.
Then, hybridized eigenmodes $\omega_\pm$ are calculated by solving the determinant of Eq.~(\ref{OmegaMP}).
Accordingly, the coupling strength at the original modes crossing point ($\omega_{\rm m} = \omega_{\rm c}$) is given by
\begin{align}
g \equiv \frac{\omega_+ - \omega_-}{2}\bigg|_{\omega_{\rm m} = \omega_{\rm c}}
.\label{gNRWA}
\end{align}

In this paper, we assume circular cross sections for both the inductor and the ferromagnet, characterized by $d$ and $d_{\rm M}$ as the diameters of the inductor and ferromagnet, respectively. Then, we have the volume ratio factor $\eta_{\rm S} = d_{\rm M}^2/d^2$. We use circuit parameters: $L = 6.2$~nH (for $N = 5$, $l = 15$~mm, and $d = 2$~mm), $C = 1$~pF, and $R = 1$~$\Omega$, which corresponds to $k_{\rm c} = 0.947$, $\omega_{\rm c}/(2\pi) = 2.0$~GHz, and $\beta = 6.3\times10^{-3}$. For simplicity, $k_{\rm m} = k_{\rm c}$ is assumed. 
By assuming ${\rm Y_3Fe_5O_{12}}$, we also use material parameters: $\gamma = 1.76\times10^{11}$~${\rm T^{-1}s^{-1}}$, $\alpha = 10^{-4}$, and $M_{\rm s} = 1.6\times10^5$~Am$^{-1}$ \cite{Mita25PRAP,Bai15}. 

To investigate how the system damping ($\alpha_{\rm m}$, $\beta$) affect the hybridized MP modes \cite{Chiba25}, we discuss eigenfrequencies of the MP modes for $d_{\rm M}/d = 0.02$ and $d_{\rm M}/d = 1$, which are displayed in Fig.~\ref{Fig:system}~(c). For $d_{\rm M}/d = 0.02$, we have coupling ratios: $g/\omega_{\rm c} = 0.01$ at $H_{01}$ and $g/\omega_{\rm c} = 0.02$ at $H_{02}$.
Here, $H_{01} = 70$~mT and $H_{02} = 138$~mT are values of the external magnetic field at each original mode crossing point, which are shown in Fig.~\ref{Fig:system}~(c). At these two points, the condition of $g/\omega_{\rm c} > \alpha_{\rm m},\beta$ but $g/\omega_{\rm c} < 0.1$ indicate that the MPs for $d_{\rm M}/d = 0.02$ reside in the SC regime. In contrast, for $d_{\rm M}/d = 1$, we have coupling ratios: $g/\omega_{\rm c} = 0.65$ at $H_{01}$ and $g/\omega_{\rm c} = 1.1$ at $H_{02}$, indicating that they reach the NSC regime.

\section{Magnonic Kerr nonlinearity and Duffing model}

Here, we discuss the nonlinear effects on the dynamics of MPs. To elucidate the origin of these effects, we first consider an extreme case of the magnon dynamics: a very weak external field, $H_0 \ll M_{\rm s}/2$, is applied to the $z$ direction in Fig.~\ref{Fig:system}~(a). In this limit, the magnetic anisotropy (demagnetization field) term in Eq.~\eqref{Um} dominates the total magnetic energy, which can be approximated as $U_{\rm m} \approx \mu_0 M_{\rm s}^2 (1 - m_x^2)/4$. Defining the transverse magnetization (magnon) as $m_{\perp} = m_y + i m_z$ and employing a second-order approximation, $m_x \approx 1 - |m_{\perp}|^2/2$, Eq.~\eqref{LLG} can be written in the following approximate form:
\begin{align}
 \frac{dm_{\perp}}{dt}
\approx 
i\frac{\gamma\mu_0M_{\rm s}}{2}m_{\perp} - i\frac{\gamma\mu_0M_{\rm s}}{4}|m_{\perp}|^2m_{\perp}
,\label{2ndapproxLLG}
\end{align}
in which the coupling term to photons and the Gilbert damping term are not displayed for simplicity.
Equation~\eqref{2ndapproxLLG} indicates that the full LLG equation with the total magnetic energy in Eq.~\eqref{Um} exhibits features of a Duffing oscillator \cite{Strogatz15,Elyasi20,Tatsumi25}, characterized by the anharmonic force $\sim x^3$, where $x$ denotes the canonical position. The Duffing oscillator exhibits an amplitude-dependent frequency, which corresponds to the Kerr nonlinearity \cite{Zhang24van}. Accordingly, in Eq.~\eqref{2ndapproxLLG}, the nonlinear frequency shift proportional to $|m_{\perp}|^2$ can be interpreted as the magnonic analog of the Kerr nonlinearity \cite{Suzuki25}.
We note, however, that this analytical correspondence between the LLG and Duffing equations \cite{Strogatz15} is generally valid only in special cases \cite{Shukrinov21,Suzuki25}.

A global correspondence between the LLG equation and the Duffing equation at the topological level has recently been demonstrated in our previous work \cite{Tatsumi25}. This correspondence is established based on the nature of equilibrium points and the existence of homoclinic orbits, analyzed via linear stability analysis of the LLG equation. Similar to the Duffing oscillator, the presence of a pair of homoclinic orbits in the phase portraits is associated with the emergence of chaos in the magnetization dynamics \cite{Tatsumi25}.
Following the same approach, Figs.~\ref{Fig:system}~(d) and (e) summarize the global correspondence between Eq.~\eqref{LLG} and the Duffing equation in terms of phase portraits. Note that, since the Kerr nonlinearity originates purely from the magnetic energy, the coupling to photons and the Gilbert damping term are neglected. Roughly speaking, these phase portraits correspond to the isoenergetic curves of Eq.~\eqref{Um2}. Figure~\ref{Fig:system}~(d) shows that two distinct homoclinic orbits emanate from the same saddle point and enclose each center in a butterfly-like shape, indicating that the topology of the phase portrait is identical to that of the Duffing oscillator \cite{Tatsumi25}. It is worth noting that horseshoe chaos can arise when a homoclinic orbit exists in the phase space \cite{Wiggins03}.
In contrast, as seen in Fig.~\ref{Fig:system}~(e), the homoclinic orbits vanish while the phase portraits remain anisotropic due to the Kerr nonlinearity (magnetic anisotropy). We also note that the MP system described by Eqs.~\eqref{LLG} and \eqref{Maxwell} (for ${\bf H}_V(t) = {\bf 0}$) is mathematically a fourth-order autonomous dynamical system, whose phase-space topology is similar to that of the H\'{e}non-Heiles system \cite{Henon64,Kasperczuk95}.

\section{Numerical demonstration}

Since we are interested in the influence of the Kerr nonlinearity on the hybridized MP modes shown in Fig.~\ref{Fig:system}(c), we focus on the Hamiltonian dynamics of MPs by setting $\alpha = \beta = 0$ in Eqs.~\eqref{LLG} and \eqref{Maxwell}. We then compute the free oscillations of MPs at the original modes crossing points ($H_0 = H_{01}, H_{02}$) as well as at the critical field of the soft magnons ($H_0 = H_{0c}$). Without loss of generality, we consider the SC case with $d_{\rm M}/d = 0.02$ (corresponding to $g/\omega_{\rm c} \approx 0.01$) and the NSC case with $d_{\rm M}/d = 1$ ($g/\omega_{\rm c} \approx 1$).

\begin{figure}[ptb]
\begin{centering}
\includegraphics[width=0.47\textwidth,angle=0]{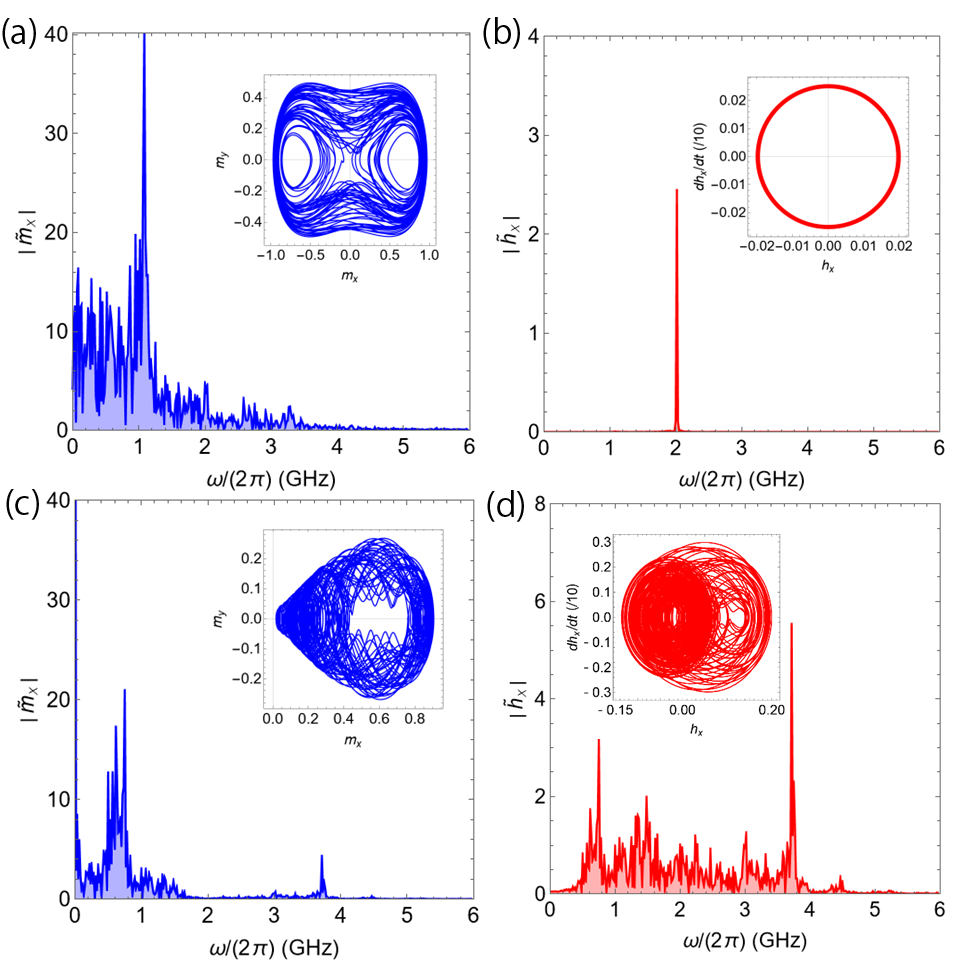} 
\par\end{centering}
\caption{Fourier spectra of (a) magnon and (b) photon dynamics at $H_0 = H_{01}$ for the SC case with $d/d_{\rm M} = 0.02$.
(c),(d) Fourier spectra for the NSC case with $d/d_{\rm M} = 1$. Insets show the corresponding phase-space trajectories of the magnon and photon dynamics. Calculations are performed with the initial conditions $h_x(0) = 0.02$ and $m_x(0) = 0$ under ${\bf H}_V(t>0) = {\bf 0}$. The Fourier spectra are obtained from the dynamics in the time window $t = 440$--500~ns.
}
\label{Fig:H01}
\end{figure}

\subsection{Mode crossing point: $H_0 = H_{01}$}

We compute the dynamics of MPs at $H_0 = H_{01}$, corresponding to the phase portraits containing the homoclinic orbit [see Fig.~\ref{Fig:system}~(d)]. Figures~\ref{Fig:H01}~(a) and (b) show the Fourier spectra of the magnon ($m_x$) and photon ($h_x$) dynamics for the SC case with $d/d_{\rm M} = 0.02$, where $m_x$ and $h_x$ denote the $x$-components of ${\bf m}(t) = {\bf M}(t)/M_{\rm s}$ and ${\bf h}(t) = {\bf H}(t)/M_{\rm s}$, respectively. As seen, the Fourier spectrum of magnons is broad and the magnetization trajectory covers a wide area in phase space [see inset]. Moreover, the magnon dynamics exhibits a strange-attractor-like trajectory, closely resembling the chaotic magnetization dynamics reported in our previous work \cite{Tatsumi25}. In contrast, the Fourier spectrum of photons shows an almost single peak and the Rabi-like splitting in Fig.~\ref{Fig:system}~(c) appears to be suppressed by the Kerr nonlinearity \cite{Suzuki25}, with the corresponding trajectory remaining closed (periodic) in phase space [see inset]. In addition, for $d/d_{\rm M} = 0.02$, the photon system volume is much larger than that of magnons so that the photon dynamics is only weakly affected by the magnon dynamics. These observations indicate that the MPs are effectively decoupled. Then, we can treat the MP as a unidirectional coupling scheme \cite{Gui24} and effectively identify the current system with the third-order autonomous dynamical system composed of driven magnons \cite{Tatsumi25}.

Figures~\ref{Fig:H01}~(c) and (d) show the Fourier spectra of magnon and photon dynamics for the NSC case with $d/d_{\rm M} = 1$. Both spectra are broad although the Rabi-like splitting in Fig.~\ref{Fig:system}~(c) is still slightly visible. This result is in sharp contrast to the SC case, indicating that MPs persist in the NSC regime even in the presence of the Kerr nonlinearity. Moreover, the photon dynamics exhibits a strange-attractor-like trajectory covering a wide area in phase space, which may have potential applications in secure communication. These observations indicate that the system is a four-order autonomous dynamical system composed of MPs. Unlike our previous work \cite{Tatsumi25} (or the SC case above) involving third-order chaos, the additional dimensionality due to the photon system may complicate the mathematical interpretation of chaos, potentially leading to hyperchaos \cite{Tanekou23}, which is beyond the scope of this paper.

\begin{figure}[ptb]
\begin{centering}
\includegraphics[width=0.47\textwidth,angle=0]{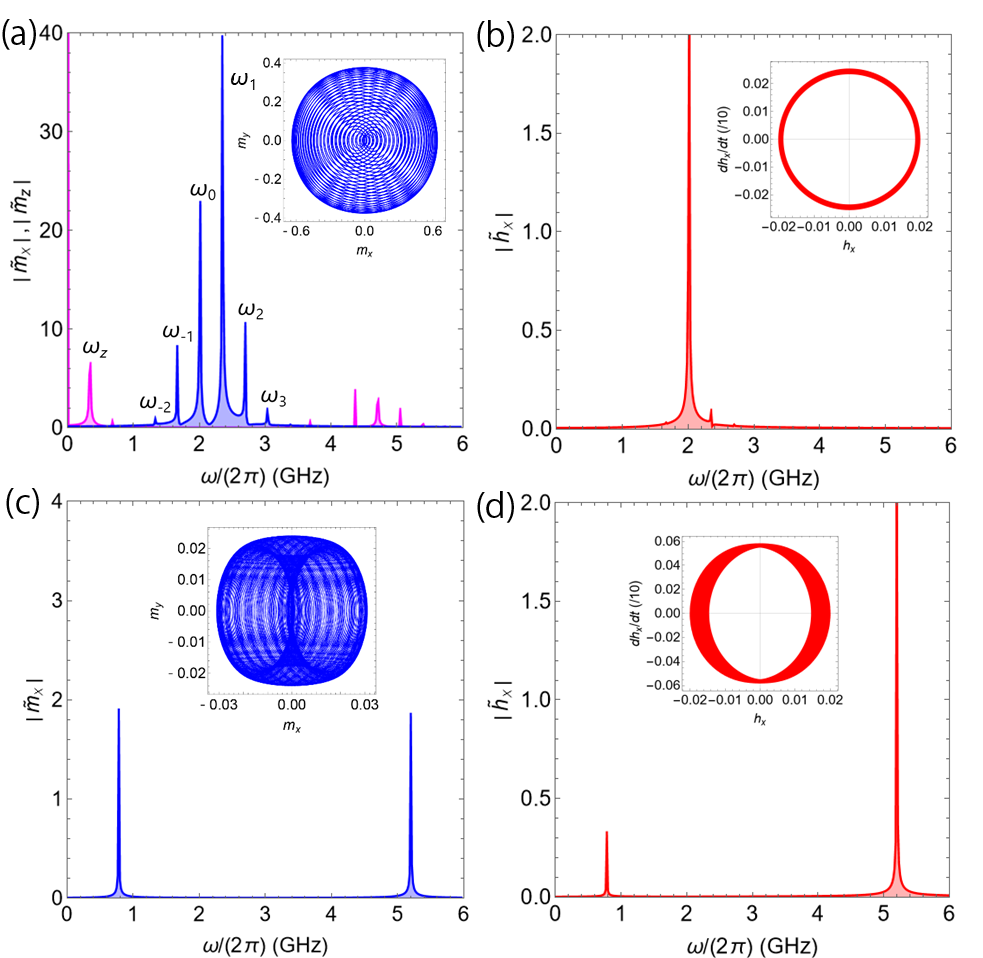} 
\par\end{centering}
\caption{Fourier spectra of (a) magnon and (b) photon dynamics at $H_0 = H_{02}$ for the SC case with $d/d_{\rm M} = 0.02$. The Fourier spectrum of $m_z$ is shown in magenta.
(c),(d) Fourier spectra for the NSC case with $d/d_{\rm M} = 1$. Insets show the corresponding phase-space trajectories of the magnon and photon dynamics. Calculations are performed under the same conditions as in Fig.~\ref{Fig:H01}.
}
\label{Fig:H02}
\end{figure}

\subsection{Mode crossing point: $H_0 = H_{02}$}

We compute the dynamics of MPs at $H_0 = H_{02}$, corresponding to the phase portraits in Fig.~\ref{Fig:system}~(e). Figures~\ref{Fig:H02}~(a) and (b) show the Fourier spectra of magnon and photon dynamics for the SC case with $d/d_{\rm M} = 0.02$. As seen, the Fourier spectrum of magnons ($m_x$) exhibits comb-like frequency sidebands. In contrast, similar to the case of $H_0 = H_{01}$, the Fourier spectrum of photons shows an almost single peak and the Rabi-like splitting is absent.
To facilitate the discussion of frequency-sideband generation, we label the positions of the main mode ($\omega_0$) and the sidebands by $\omega_n$ ($n = -2, \dots, 3$), as shown in Fig.~\ref{Fig:H02}~(a). The frequency differences are defined as $\Delta_n/(2\pi) = \omega_{n+1}/(2\pi) - \omega_n/(2\pi)$. These values satisfy $\Delta_{-2}/(2\pi) = \dots = \Delta_2/(2\pi) \approx 0.34~\mathrm{GHz}$, which corresponds to the main peak ($\omega_z/(2\pi) = 0.34~\mathrm{GHz}$) of the Fourier spectrum of $m_z$ arising from the Kerr nonlinearity \cite{Suzuki25}. Thus, the frequency sidebands can be interpreted as resulting from sum- and difference-frequency generation (superposition) between the main mode ($\omega_0 \approx \omega_{\rm m}$) and the nonlinear mode ($\omega_z$), expressed as
\begin{align}
\omega_{n} = \omega_{0} + n\omega_{z}~(n = -2,\cdots,3)
.\label{Comb}
\end{align}

\begin{figure}[ptb]
\begin{centering}
\includegraphics[width=0.47\textwidth,angle=0]{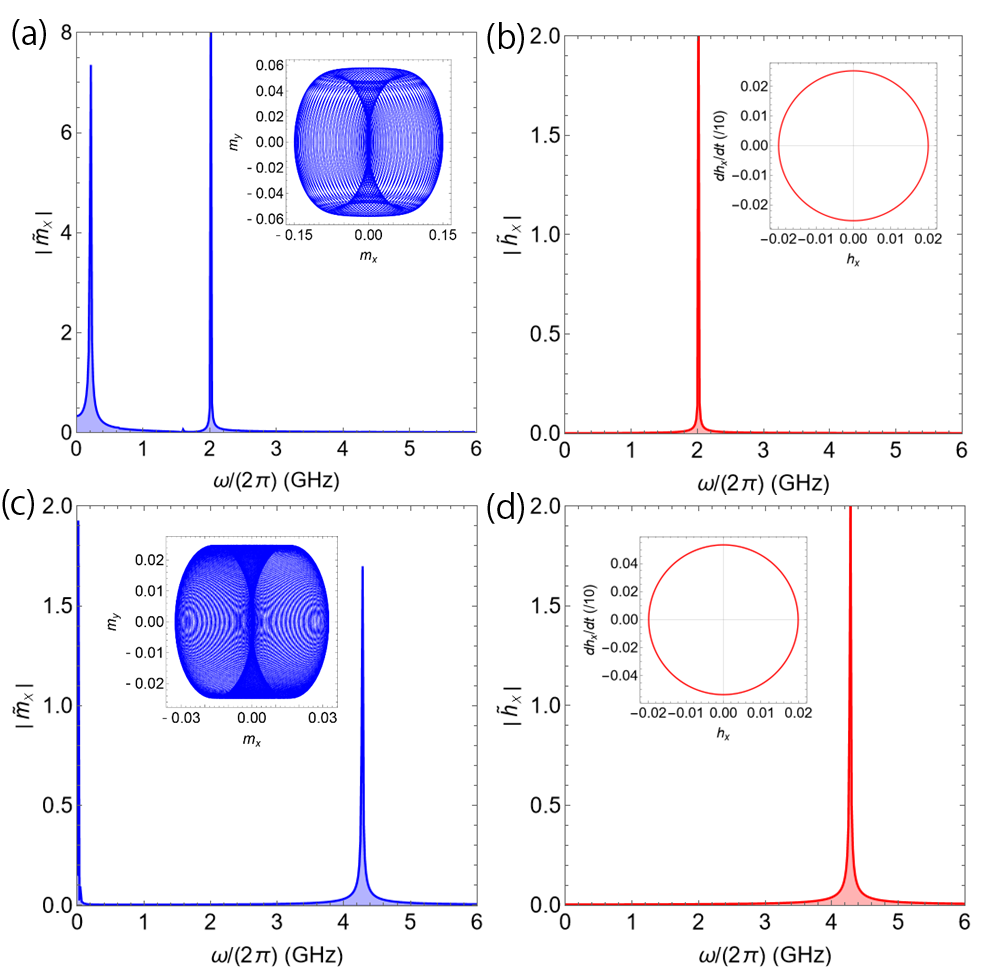} 
\par\end{centering}
\caption{Fourier spectra of (a) magnon and (b) photon dynamics at $H_0 = H_{0c}$ for the SC case with $d/d_{\rm M} = 0.02$.
(c),(d) Fourier spectra for the NSC case with $d/d_{\rm M} = 1$. Insets show the corresponding phase-space trajectories of the magnon and photon dynamics. Calculations are performed under the same conditions as in Fig.~\ref{Fig:H01}.
}
\label{Fig:H0c}
\end{figure}

Figures~\ref{Fig:H02}~(c) and (d) show the Fourier spectra of magnon and photon dynamics for the NSC case with $d/d_{\rm M} = 1$. As seen, the Rabi-like splitting is clearly visible in both spectra. The magnon dynamics exhibits quasi-periodic behavior. This result is in sharp contrast to the SC case, indicating that MPs robustly persist in the NSC regime.

\subsection{Critical field point: $H_0 = H_{\rm 0c}$}

We compute the dynamics of MPs at the critical field corresponding to the soft-mode ($H_0 = H_{\rm 0c}$). Figures~\ref{Fig:H02}~(a) and (b) show the Fourier spectra of magnon and photon dynamics for the SC case with $d/d_{\rm M} = 0.02$. As seen, a finite excitation gap appears in the soft-mode of the magnon spectrum. This gap arises from the confinement effect due to the higher-order magnon potential, which is the origin of the Kerr nonlinearity.

Figures~\ref{Fig:H0c}~(c) and (d) show the Fourier spectra for the NSC case with $d/d_{\rm M} = 1$. The finite excitation gap in the soft-mode is suppressed. Furthermore, photons couple to the soft-mode magnons, resulting in Rabi-like splitting, in sharp contrast to the SC case. These observations indicate that MPs robustly persist in the NSC regime. In other words, the NSC regime overcomes the magnonic Kerr nonlinearity, which may support divergence-like behavior of quantum squeezing by exploiting the soft (zero)-mode magnons \cite{Lee23,Silaev23} beyond the linear spin-wave approximation.

\section{Summary}

In summary, we have theoretically investigated the influence of magnonic Kerr nonlinearity on hybridized MP modes with a soft-mode in easy-axis ferromagnets. Based on an effective circuit model capable of describing the NSC regime, we computed the dynamics of MPs at the original modes crossing points as well as at the critical field of soft magnons.
We demonstrated that MPs exhibit chaotic and frequency-comb-like behaviors at the original modes crossing points in the typical SC regime ($g/\omega_{\rm c(m)} \approx 0.01$). We also found that a finite excitation gap emerges in the soft-mode due to the Kerr nonlinearity, particularly in the SC regime. In contrast, such nonlinear behaviors are strongly suppressed in the NSC regime ($g/\omega_{\rm c(m)} \approx 1$). This suppression supports the validity of the linear spin-wave approximation, which is essential for quantum squeezing using soft (zero)-mode magnons \cite{Lee23,Silaev23}.

\begin{acknowledgments}
The author thanks R. Tatsumi, R. Suzuki, H. Chiba, T. Taniguch, and T. Oto for valuable discussions. This work was supported by Grants-in-Aid for Scientific research (Grants No.~22K14591 and No.~25H02105). 
\end{acknowledgments} 

\section*{Data Availability}

The data that support the findings of this study are available from the corresponding author upon reasonable request.




\end{document}